\documentclass[aps,pre,onecolumn,10pt]{revtex4}
\usepackage{amsfonts,amscd}
\usepackage{amssymb}
\usepackage{mathtools}
\usepackage{amsmath}
\usepackage{amscd}
\usepackage{amsmath}
\usepackage{lineno,hyperref}
\modulolinenumbers[5]
\usepackage{float}
\usepackage{xcolor}
\usepackage[]{graphicx}
\usepackage{subfigure} 
\usepackage{chngcntr}       
\begin{document}        
\title{Amplitude chimeras and bump states with and without frequency entanglement:\\ a toy model}     
\author{A. Provata}                           
\address{Institute of Nanoscience and Nanotechnology,     
National Center for Scientific Research ``Demokritos'', GR-15341 Athens, Greece                     
}     
\date{\today}             

\begin{abstract}

When chaotic oscillators are coupled in complex networks a number of interesting synchronization phenomena emerge.
Notable examples are the frequency and amplitude chimeras, chimera death states, solitary states as well as combinations
of these. In a previous study [Journal of Physics: Complexity, 2020, 1(2), 025006], a toy model  was introduced 
addressing possible mechanisms behind the formation of frequency chimera states. In the present study a variation
of the toy model is proposed to address the formation of amplitude chimeras. The proposed oscillatory model is now
equipped with an additional 3rd order equation modulating the amplitude of the network oscillators. This way, 
the single oscillators are constructed as bistable in amplitude and depending
on the initial conditions their amplitude may result in one of the two stable fixed points.
Numerical simulations demonstrate that when these oscillators are nonlocally coupled in networks,
they organize in domains with alternating amplitudes (related to the two fixed points), naturally forming amplitude chimeras.
A second extension of this model incorporates nonlinear terms merging amplitude together with frequency, 
and this extension allows for the spontaneous production of composite amplitude-and-frequency
chimeras occurring simultaneously in the network. 
Moreover the extended model allows to understand the emergence of 
bump states via the continuous passage from chimera states, when both fixed point amplitudes are positive, 
to bump states when one of the two fixed points vanishes. 
The synchronization properties of the network are studied 
as a function of the system parameters for the case of amplitude chimeras, bump states and  
composite amplitude-and-frequency chimeras. 
The proposed mechanisms of creating domains with
variable amplitudes and/or frequencies provide a generic scenario for understanding
  the formation of the complex synchronization phenomena
observed in networks of coupled nonlinear and chaotic oscillators.
\end{abstract}
 
\maketitle
\noindent{\bf Keywords:} {Coupled oscillator networks; bistability; double-well synchronization; amplitude chimeras; frequency chimeras; entanglement;
amplitude death states; oscillations death; bump states.}


\everymath{\displaystyle}

\section{Introduction}  
\label{sec:intro}


Chimera states have been on the focus of research in the areas of nonlinear dynamics and complex systems 
during the past two decades \cite{kuramoto:2002,kuramoto:2002a,abrams:2004,panaggio:2015,schoell:2016,zheng:2016,oomelchenko:2018}.
In a network of coupled nonlinear oscillators chimera states may emerge in specific parameter regions, 
leading to the splitting of the network into multiple coexisting coherent and incoherent domains \cite{kuramoto:2002}.
In the original studies chimera states were reported in systems of identical oscillators with nonlocal connectivity,
while later studies have demonstrated emerging chimeras under local, nonlocal or global connectivity schemes with delays
and/or with non-identical elements \cite{zakharova:2020}. A important characteristic of the typical chimera state was the frequency deviation
in the incoherent regions: while all elements of the coherent regions have the same frequency of
oscillations, the frequencies vary drastically in the incoherent regions \cite{kuramoto:2002,provata:2020,rontogiannis:2021}.

\par Besides frequency variations, recent studies of different dynamical systems beyond stable nonlinear oscillators (e.g., limit cycles) have demonstrated 
that amplitude variations may also occur in coupled networks, especially in the case of chaotic oscillators. In particular, coupled
Lorenz systems or other chaotic oscillators  lead to   chimera states with variable amplitude and frequency
\cite{shepelev:2017,shepelev:2018,bogomolov:2017,muni:2020,tsakalos:2022}. These studies have lead to the search for chimera states 
with pure amplitude variations and are now known under the terms: pure amplitude chimeras, amplitude-mediated chimeras,
amplitude-death chimeras and amplitude-modulated chimeras \cite{zakharova:2016,sethia:2014,banerjee:2018,sathiyadevi:2018,kundu:2018,bi:2022}.
The interplay of amplitude and frequency variations adds an extra dimension to the complexity of the activity in
systems of coupled nonlinear oscillators. 

\par Bump states are localized states of activity also found in networks of coupled elements. In bump states incoherent domains with gradually varying frequencies
are surrounded by locked coherent domains \cite{oomelchenko:2021}. These states are different from chimera states. In both chimera and bump states
 we have the presence of one
or multiple incoherent domains, but in chimera states the coherent domains oscillate coherently while in the bump states they are locked
 \cite{oomelchenko:2021,laing:2020,laing:2021,rontogiannis:2021,tsigkri:2017}.
Bump states appear as a self-organizing phenomenon in networks of coupled nonlinear oscillators and they are a result of synergy between the various units. 
They are commonly discussed in ensembles of coupled neurons (neuroscience) but also in the context of other fields, such as physics, chemistry, and biology.

\par States involving complex synchronization patterns and chimera states have been experimentally realized in  networks
of coupled mechanical oscillators \cite{martens:2013,blaha:2016,dudkowski:2016},
chemical oscillators \cite{tinsley:2012,nkomo:2013,taylor:2015},
electronic oscillators \cite{gambuzza:2014,gambuzza:2020,rosin:2014,english:2017},
 nonlinear optics \cite{hagerstrom:2012}, 
in electrochemical applications \cite{wickramasinghe:2013,wiehl:2021,patzauer:2021}  
and in laser physics \cite{uy:2019}.

\par Motivated by the discovery of amplitude chimeras in networks of chaotic oscillators, we now seek the
mechanisms that drive the coexistence of regions with different amplitudes even if all oscillators are
identical and identically coupled. In a previous study, a toy model was devised to
include frequency bifurcations \cite{provata:2020}. As a result of the frequency bifurcations, together
with the cooperation between elements in a broad nonlocal neighborhood, alternating
regions with two different frequencies develop. 
Within each region all elements have constant common frequency, but the frequency in successive regions
is different. These regions constitute the coherent domains of the network.
To bridge the frequency gap between consecutive coherent domains, intermediate regions of variable
frequency are developed which serve as transition regions and are characterized
by a gradient of frequencies.
These transition regions constitute the incoherent domains of the network \cite{provata:2020}. 
Applying these ideas to amplitudes instead of frequencies, we here
modify the system in such a way that now bifurcation takes place not in the frequency of the oscillators
but in the amplitude. This bifurcation will give rise to regions with at least two different amplitudes
of oscillations and variable amplitudes in the joining domains. This way amplitude chimeras will be
developed as will be presented in the next sections. 
\par As an additional extension, we complexify further the nodal dynamics by entangling the
constant frequency of the elements with the variable amplitude dynamics (which undergo bifurcations).
As a result of this second extension, we obtain composite global dynamics where coexisting domains
are formed with both amplitude and frequency alternations as well as bump states in particular
parameter regions.
Both proposed mechanisms extending the frequency chimeras to amplitude and to amplitude-frequency ones
are generic, and can be
at the origin of complex chimera states.

More specifically, in the previous study \cite{provata:2020}, a toy model was introduced where a nonlinear oscillator was
equipped with an additional equation giving rise to frequency bifurcations. The nonlinear toy oscillator
was described by the state variables $x$, $y$ and by the instantaneous phase velocity variable $\omega (=\omega (t))$, as follows:
\begin{subequations}
\begin{align}
\frac{dx}{dt} &= -ax+\frac{aR}{\sqrt{x^2+y^2}}x-\omega y
\label{eqno01a} \\
\frac{dy}{dt} &= -ay+\frac{aR}{\sqrt{x^2+y^2}}y+ \omega x 
\label{eqno01b} \\
\frac{d\omega}{dt} &= c_\omega (\omega -\omega_l)(\omega -\omega_c)(\omega -\omega_h)
\label{eqno01c}
\end{align}
\label{eqno01}
\end{subequations}
In Eqs.~\eqref{eqno01a} and~\eqref{eqno01b}, $R$ denotes the final amplitude of the oscillations and $a$ is the
relaxation exponent.
The value of the constant $c_\omega$ in Eq.~\eqref{eqno01c} governs the frequency of the nonlinear oscillator.
As in previous studies, in the following we use the terms 
``frequency'', or ``phase velocity''
or ``angular frequency'', interchangeably,  to refer to the values of $\omega$, although the angular frequency
is related to the frequency $f$ by a factor of $2\pi$, $\omega =2\pi f$. 
The constants $\omega_l$ (standing for $\omega_{\rm low}$), $\omega_c$ (standing for $\omega_{\rm center}$)
and $\omega_h$ (standing for $\omega_{\rm high}$) denote the three fixed points of Eq.~\eqref{eqno01c}
and without loss of generality we assume that $\omega_l < \omega_{\rm c} < \omega_h$. The stability properties
of the three fixed points is of great importance for the dynamics of the system.

\begin{enumerate}
\item If $c_\omega=0$, then the value of $\omega$ is constant; Eqs.~\eqref{eqno01a} and
\eqref{eqno01b} can be explicitly solved and the oscillator presents exponential relaxation to a circle with radius $R$, 
with relaxation exponent $a$. The solution is:
\begin{subequations}
\begin{align}
x(t) &= R(1-Ae^{-at}) \cos (\omega t)\\
y(t) &= R(1-Ae^{-at}) \sin (\omega t).
\end{align}
\label{eqno02}
\end{subequations}
\noindent Here $A$ determines the
initial position inside a circle of constant radius $R$ and constant phase velocity $\omega$. If the 
oscillator starts from position $\left( x_0,y_0\right)$ then $A$ is defined by the equation: $ x_0^2+y_0^2=R^2(1-A)^2$. 
As time increases the term $Ae^{-at}$ decreases exponentially to $0$, giving rise to a purely circular, stable orbit
of radius $R$ and constant $\omega$.
\item If $c_\omega > 0$ is positive, then $\omega =\omega (t)$ becomes time dependent.
 Eq.~\eqref{eqno01c} has three fixed points, $\omega =\omega_l , \> \omega_c $ and $\omega_h$.
From the point of view of stability, $\omega_c$ is an attractor and $\omega_l$ and $\omega_h$ are repellers. 
Starting with an arbitrary initial $\omega (t=0)$-value the system can only end-up with final $\omega =\omega_c$.
\item Of greater interest is the case where $c_\omega < 0$ is negative. Here also $\omega =\omega (t)$ is time dependent.
Stability-wise,  $\omega_c$ is a repeller
 and $\omega_l$ and $\omega_h$ are attractors. The system now may exhibit frequency bistability: 
depending on the initial value $\omega (t=0)$  the system can end-up in one of the two stable frequencies,
 $\omega_l$ or $\omega_h$.
\end{enumerate}
In Ref.~\cite{provata:2020}, the last option of $c_\omega$-values was considered, $c_\omega < 0$, which allows for bistability.
It was reported that when many oscillators described by Eq.~\eqref{eqno01} are coupled in a network, two-level synchronization
is achieved, where the network splits in domains alternating in frequencies between the values $\omega_l$ or $\omega_h$.
Joining these domains, other incoherent domains develop to bridge the frequency gaps.  The multiple stable solutions
of Eq.~\eqref{eqno01c}
lead to the emergence of a chimera state with two-levels of synchronization
\cite{provata:2020,hizanidis:2014,dudkowski:2014,shepelev:2017,shepelev:2019}.

\par In the present study we present a variant of the above toy model, where the bifurcation scenario concerns the
amplitude of the oscillations, whereas the frequency remains constant. We present numerical evidence that 
when many such oscillators are coupled in a network, the network splits in oscillating domains of variable amplitude
with constant frequency. These composite states are the 
amplitude chimeras \cite{zakharova:2016,gjurchinovski:2017,zakharova:2020}.
Other parameter values are explored which lead to chimera amplitude death states. We also explore
 entanglement of the amplitude and 
frequency variables. The entanglement leads to complex chimera states with modulated frequencies in the network.
More complex chimera forms are also discussed in the case of entanglement, when both 
oscillators' frequencies and amplitudes  undergo bifurcations.

\par The organization of the work is as follows: 
In Sec.~\ref{sec:model0} we introduce the variant toy model, i.e., the uncoupled two-amplitude nonlinear oscillator
and briefly discuss its steady state properties. In the same section we also consider amplitude-frequency entanglement.
In Sec.~\ref{sec:coupled} we couple the toy-oscillators in a ring network and we discuss
synchronization measures. In Sec.~\ref{sec:parameters} we examine the emerging amplitude chimeras
and chimera death states as a function of the coupling variables. In particular, in Sec.~\ref{sec:results-without-entanglement}  we consider the non-entangled 
case and  in Sec.~\ref{sec:results-with-entanglement}  we present the
extension of the toy model, where amplitude and frequency variables are entangled and we 
discuss the influence of the amplitude bifurcations in the frequency domain.
In Sec.~\ref{sec:results-for-bump-states} we discuss the conditions under which the bump states emerge.
In the Concluding section we recapitulate our main results and discuss open problems.

\section{The uncoupled two-amplitude oscillator}
\label{sec:model0}

In this section, we first introduce the variant toy model,
where the modulation now affects the amplitude of the oscillations and not the frequency, 
in analogy with Ref.~\cite{provata:2020}. In the second part of this section, Sec.~\ref{sec:entanglement0},
we extend the variant toy model to introduce an elemental entanglement between
frequency and amplitude in the oscillatory part of the dynamics. This way, the amplitude modulations
affect the frequency of the motion which varies nonlinearly in time.
All results in the present section concern the dynamics of uncoupled oscillatory units;
the effects of coupling will be discussed in Secs.~\ref{sec:coupled}, ~\ref{sec:parameters} and ~\ref{sec:results-for-bump-states}.

\subsection{Two-amplitude oscillator without entanglement}
\label{sec:noentanglement0}
Starting from the toy model discussed in the Introduction and in
Refs.~\cite{provata:2020} and ~\cite{provata:2018},
we now consider the case where the frequency $\omega $ is constant, but the amplitude $r$ is time dependent, ${r(t)} $, and
is governed by bistable dynamics, as follows:
\begin{subequations}
\begin{align}
\frac{dx}{dt} &= -ax+\frac{ar}{\sqrt{x^2+y^2}}x-\boxed{\omega} y
\label{eqno03a} \\
\frac{dy}{dt} &= -ay+\frac{ar}{\sqrt{x^2+y^2}}y+ \boxed{\omega} x 
\label{eqno03b} \\
\frac{d r}{dt} &= c_r (r -r_l)(r -r_c)(r -r_h)
\label{eqno03c}
\end{align}
\label{eqno03}
\end{subequations}

\noindent The variable $\omega$ is boxed in Eqs.~\eqref{eqno03}a,b for highlighting purposes and for
comparison with Eqs.~\eqref{eqno06}a,b, later on. The time dependent amplitude is here represented by small $r$, 
instead of capital $R$ used  in the Introduction and in Refs.~\cite{provata:2018} and ~\cite{provata:2020},
where it refers to constant amplitude oscillations. Similarly, the constant $c_r$ in Eq.~\eqref{eqno03c} 
governing the amplitude dynamics takes the subscript $r$ to 
stress that it is related to the amplitude $r$ modulations. 

 \par When $c_r=0$, the amplitude takes a constant value $r(t)=R$=constant and the model reduces to
\begin{subequations}
\begin{align}
\frac{dx}{dt} &= -ax+\frac{aR}{\sqrt{x^2+y^2}}x-\omega y\\
\frac{dy}{dt} &= -ay+\frac{aR}{\sqrt{x^2+y^2}}y+ \omega x .
\end{align}
\label{eqno04}
\end{subequations}
\noindent The solution is analytically
obtained as Eqs.~\eqref{eqno05}, see also Ref.~\cite{provata:2020}. 

\begin{subequations}
\begin{align}
x(t) &= R(1-Ae^{-at}) \cos (\omega t)\\
y(t) &= R(1-Ae^{-at}) \sin (\omega t).
\end{align}
\label{eqno05}
\end{subequations}
\noindent Starting from any initial state determined as $x^2 +y^2=R^2(1-A)^2$ with $0\le A <1$, the trajectory relaxes
exponentially to a circle of constant radius $R$ and constant phase velocity $\omega$.  The relaxation exponent is
denoted by $a$. 

\par When $c_r > 0$, then $r =r (t)$ becomes time dependent.
 Eq.~\eqref{eqno03c} has three fixed points, $r =r_l , \> r_c $ and $r_h$. The subscripts $l$, $c$ and $h$ denote
the low, center and high amplitude fixed points, respectively.
From the point of view of stability, $r_c$ is an attractor and $r_l$ and $r_h$ are repellers. 
Starting with an arbitrary initial $r(t=0)$-value the system can only end-up with final amplitude $r =r_c$.
\par If $c_r < 0$ the amplitude becomes also time dependent, $r =r(t)$, but,
stability-wise, the central fixed point $r_c$ is a repeller and
 $r_l$ and $r_h$ are attractors. The system now may exhibit \underline{amplitude bistability}.
Depending on the initial amplitude $r(t=0)$  the system can end-up in one of the two stable fixed points,
 $r_l$ or $r_h$.

\par As an example, in Fig.~\ref{fig01}a we present the temporal evolution of Eqs.~\eqref{eqno03} starting
from two different initial conditions, called I (black lines)
and II (red lines). Other parameters are exactly the same: $a=1.0$, $\omega =1.0$,
$r_l=2.0$, $r_c=3.0$ and $r_h=4.0$. In panel a) we note that the two trajectories have equal frequencies but 
different amplitudes. The equality of the frequencies is numerically verified and shown in plot Fig.~\ref{fig01}b. 

\par This last option, $c_r < 0$, will be used in the next section, Sec.~\ref{sec:coupled}, when many two-amplitude
oscillators will be coupled in a ring. We will numerically demonstrate that
when the different coupled oscillators start with different, random, initial amplitudes there will be a split of the
network in domains of  alternating low $r_l$ and high $r_h$ amplitudes.

\par Before introducing the coupled system, we present in Sec.~\ref{sec:entanglement0} an extension of the amplitude
bistable toy model where amplitude $r$ and frequency $\omega$ are entangled and, as a result, they influence each other
and complexify the overall motion.

\subsection{Two-amplitude oscillator with amplitude-frequency entanglement}
\label{sec:entanglement0}

For the amplitude-frequency entanglement we consider the following extension of the 
amplitude modulation toy model:

\begin{subequations}
\begin{align}
\frac{dx}{dt} &= -ax+\frac{ar}{\sqrt{x^2+y^2}}x-\boxed{r \omega} y
\label{eqno06a} \\
\frac{dy}{dt} &= -ay+\frac{ar}{\sqrt{x^2+y^2}}y+ \boxed{r \omega } x 
\label{eqno06b} \\
\frac{d r}{dt} &= c_r (r -r_l)(r -r_c)(r -r_h)
\label{eqno06c}
\end{align}
\label{eqno06}
\end{subequations}

Namely, we introduce a factor of $r$
multiplying the constant phase velocity $\omega$ in
the last terms of Eqs.~\eqref{eqno06a} and ~\eqref{eqno06b} 
(see boxed expressions and compare with the case without entanglement
presented in Eqs.~\eqref{eqno03a} and ~\eqref{eqno03b}). This is the only differences between 
systems Eqs.~\eqref{eqno03} and ~\eqref{eqno06}.
The solutions for the variables $x$ and $y$ can be given analytically in terms of $r$ and $t$
\begin{subequations}
\begin{align}
x(t) &= r(1-Ae^{-at}) \cos (r \omega t)\\
y(t) &= r(1-Ae^{-at}) \sin (r \omega t),
\end{align}
\label{eqno07}
\end{subequations}
\noindent while the evolution of $r$ is governed by the last, independent equation~\eqref{eqno06c}.
As previously described, if $c_r <0$, the last equation~\eqref{eqno06c} has two attracting values for 
the amplitude $r$, namely, $r_l$ and $r_h$. Depending on the initial conditions $r$ may reach any of the two 
fixed points
and then the state variables $x$ and $y$ will perform oscillations with variable phase velocity, equal to $r\omega$.
This way the bifurcation of the $r$-variable in Eq.~\eqref{eqno06c}
 influences not only the amplitude of the oscillations but also modifies its frequency.

\par
\begin{figure}[t]
\includegraphics[clip,width=0.45\linewidth,angle=0]{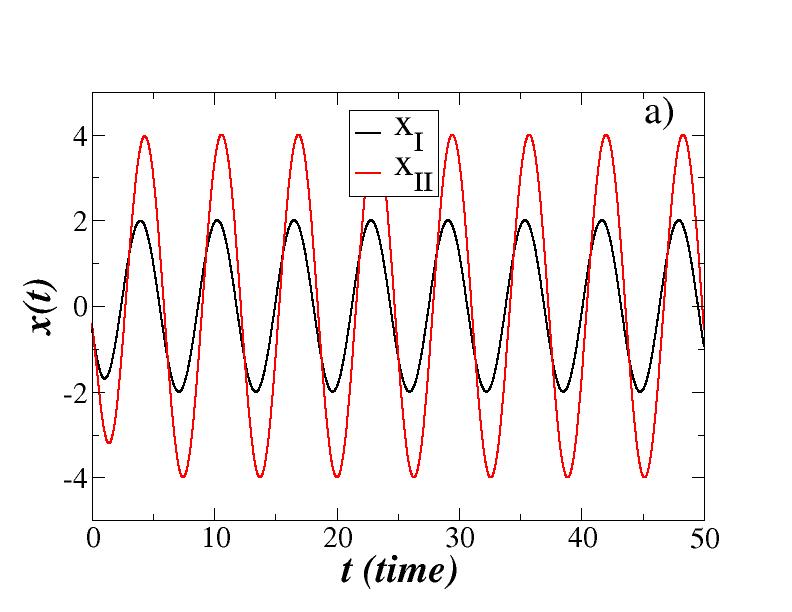}
\includegraphics[clip,width=0.45\linewidth,angle=0]{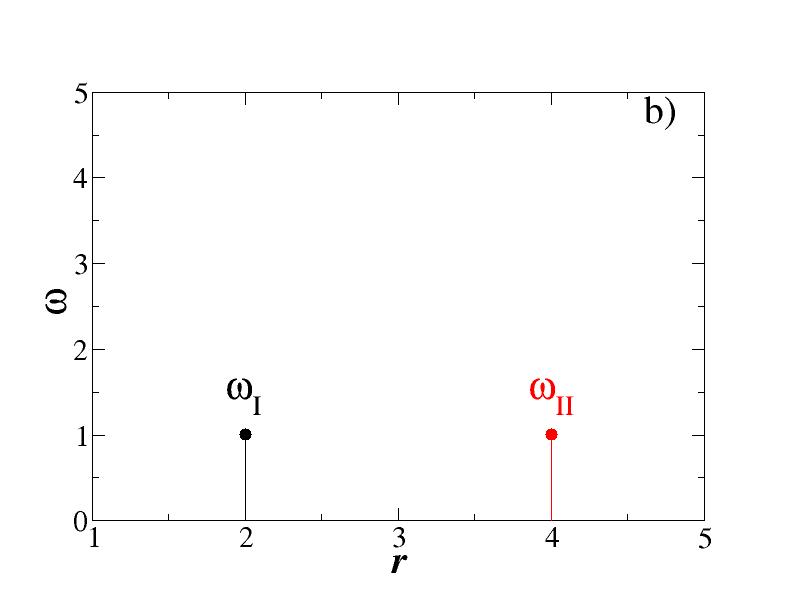}
\includegraphics[clip,width=0.45\linewidth,angle=0]{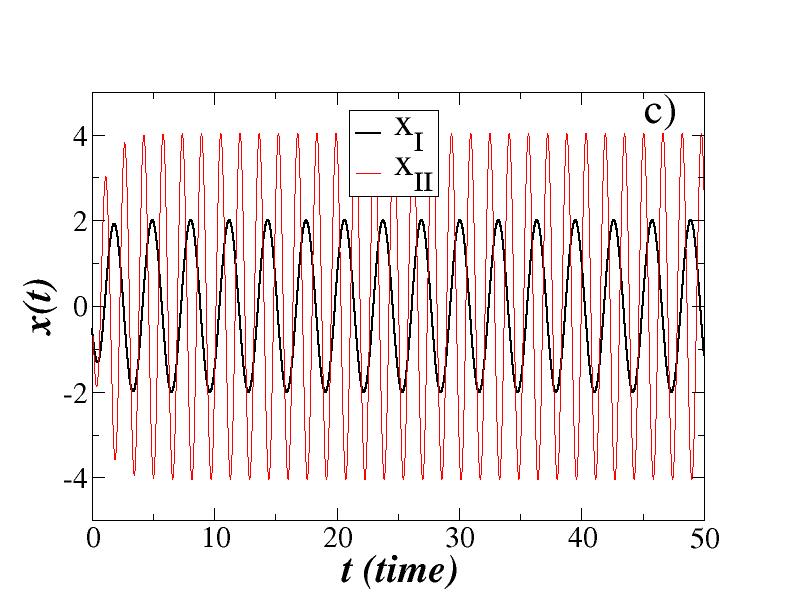}
\includegraphics[clip,width=0.45\linewidth,angle=0]{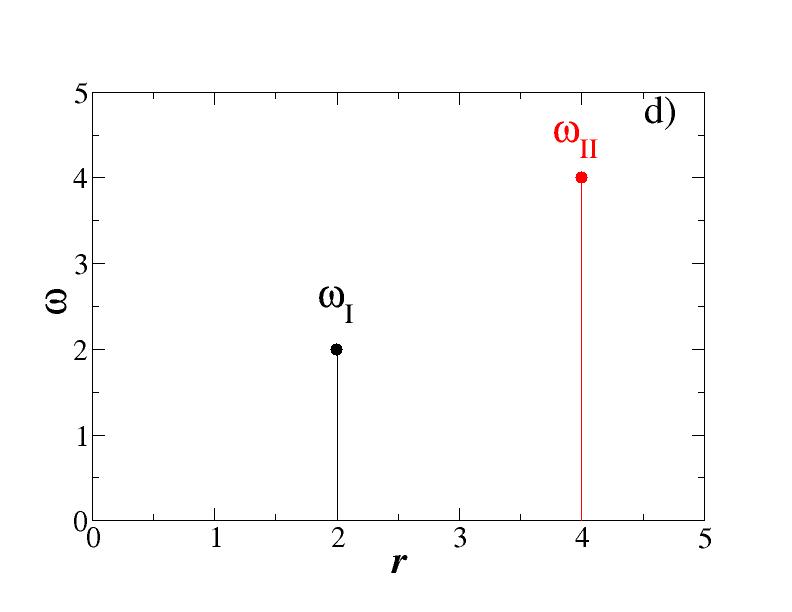}
\caption{\label{fig01} (Color online)
The uncoupled two amplitude model. The black (red) lines and dots correspond
to initial conditions I (II): a) Without entanglement (Eqs.~\eqref{eqno03}): temporal evolution of
$x_I$ and $x_{II}$  variables starting from two different initial conditions (I and II),
b) the corresponding computed mean phase velocities $\omega_I$ and $\omega_{II}$ are equal.  
c) With entanglement (Eqs.~\eqref{eqno06}): temporal evolution of $x_I$ and $x_{II}$ 
variables starting from the same initial conditions I and II as in panel a, and d) 
the corresponding computed, asymptotic mean phase velocities $\omega_I$ and $\omega_{II}$ are different.
Parameters here are: $r_l=2$, $r_c=3$, $r_h=4$, $\omega =1.0$ and $a=1.0$.
}
\end{figure}

 To compare with the case without entanglement, in Fig.~\ref{fig01}c we present the evolution
of the $x-$variable starting from precisely the same initial conditions and parameters
as in Fig.~\ref{fig01}a, but with the
entanglement as described in Eqs.~\eqref{eqno06}. We can now notice that not only the amplitude of
the oscillations but also the frequencies change as a result of the entanglement. The frequency change 
 in the two trajectories is further verified by numerically computing the 
corresponding, asymptotic, mean phase velocities, see Fig.~\ref{fig01}d.
\par The amplitude-frequency entanglement may be a source of false interpretation when observing different
frequencies in a network, as will be discussed in the next section. In such cases the
 frequency, by mistake, may be considered as a variable in the system,
while in reality variability comes from the entanglement with the amplitude.

\section{Two-amplitude oscillators coupled in a ring network}
\label{sec:coupled}

In this section we first present in Sec.~\ref{sec:coupled-without-entanglement} the case of coupled two-amplitude oscillators 
without entanglement, while in Sec.~\ref{sec:coupled-with-entanglement} the 
coupled oscillators are subjected to amplitude-frequency entanglement.
For both cases a simple ring network geometry is considered. 

\subsection{Coupled two-amplitude oscillators without entanglement}
\label{sec:coupled-without-entanglement}

Consider a ring network containing
$N$ nodes. We denote by $x_i$ and $y_i$ the state variables on the $i$th node and by $r_i$ its amplitude. 
We use the simplest nonlocal coupling scheme 
where each oscillator is linearly coupled to $S$ closest neighbors on its left and $S$ neighbors on its right. 
Therefore, each element is coupled to $2S$ closest neighbors, in total.
Furthermore, let us denote by small letter $s=2S/N$, the ratio of coupled elements $2S$
over  the total number of elements $N$ in the network. The parameter $s$ expresses the extent of nonlocal
connectivity in the network.
The coupling strength is denoted by $\sigma$ and is common for the $x_i$ and $y_i$ variables and maybe different, denoted by $\sigma_{r}$, in the 
$r_i$ variables. The phase velocity $\omega$, relaxation exponent $a$ and $c_r <0$ are constant and common to all oscillators.
 The coupled system dynamics reads:
\begin{subequations}
\begin{align}
\frac{dx_i}{dt} &= -ax_i+\frac{ar_i}{\sqrt{x_i^2+y_i^2}}x_i-\boxed{\omega} y_i + 
\frac{\sigma}{2S}\sum_{j=i-S}^{i+S}\left[ x_j-x_i\right]
\label{eqno08a} \\
\frac{dy_i}{dt} &= -ay_i+\frac{ar_i}{\sqrt{x_i^2+y_i^2}}y_i+\boxed{\omega} x_i + 
\frac{\sigma}{2S}\sum_{j=i-S}^{i+S}\left[ y_j-y_i\right]
\label{eqno08b} \\
\frac{dr_i}{dt} &= c_r (r_i -r_l)(r_i -r_c)(r_i -r_h)+\frac{\sigma_r}{2S}\sum_{j=i-S}^{i+S}\left[ r_j-r_i\right] .
\label{eqno08c}
\end{align}
\label{eqno08}
\end{subequations}
\noindent In Eqs.~\eqref{eqno08} all indices are taken$\mod{N}$. In the simulations,
random initial conditions are considered in all $3N$
variables $(x_i, y_i, r_i)$, $i=1, \cdots , N$. Since the frequency $\omega$ is common to all elements
and there is no explicit frequency bifurcation in the system, we do not expect to observe chimera states
with domains where different frequencies develop. On the other hand, Eqs.~\eqref{eqno08c} allows for the choice of
different amplitudes in different domains of the network, depending on the local choice of initial amplitudes.
This way amplitude chimeras may develop, where domains with different amplitudes
alternate in the network. This scenario may be on the basis of the amplitude chimeras observed
 in nonlocally coupled chaotic systems \cite{shepelev:2017,shepelev:2018,bogomolov:2017,muni:2020},
among other interesting synchronization phenomena.

\begin{figure}[h]
\includegraphics[clip,width=0.32\linewidth,angle=0]{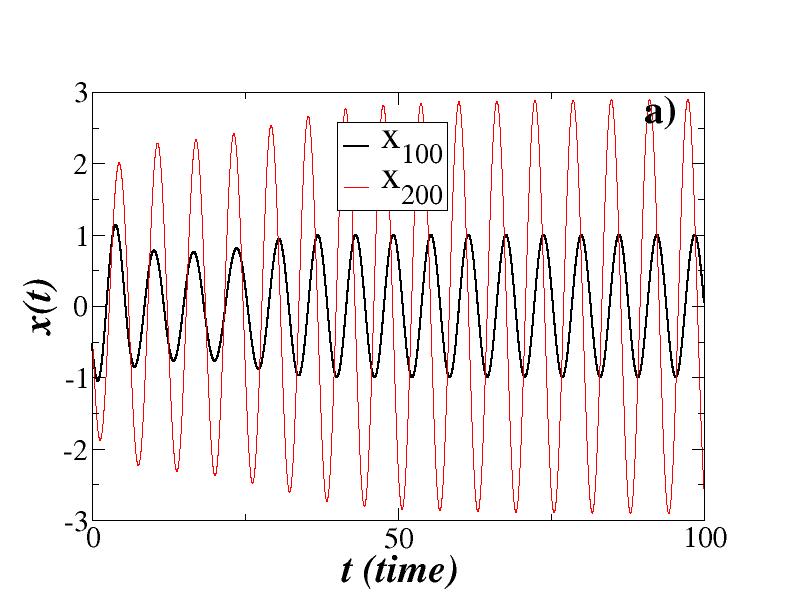}
\includegraphics[clip,width=0.32\linewidth,angle=0]{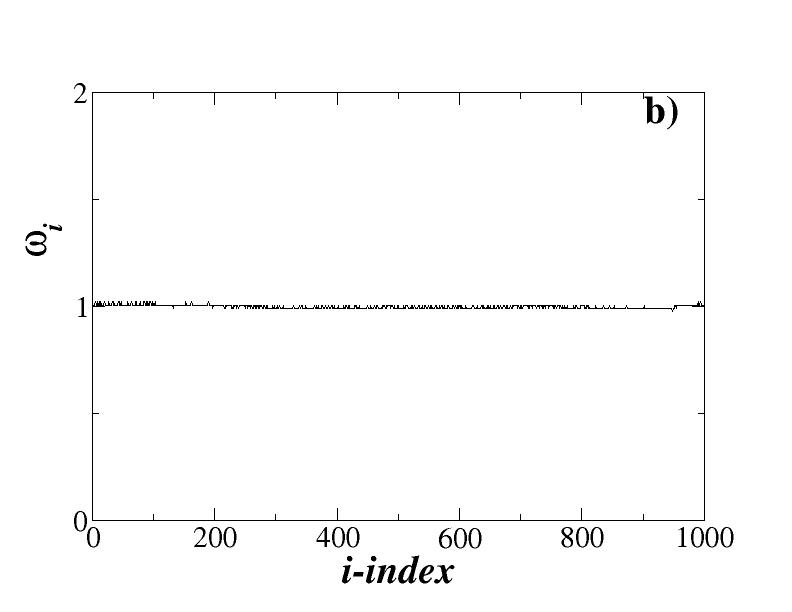}
\includegraphics[clip,width=0.32\linewidth,angle=0]{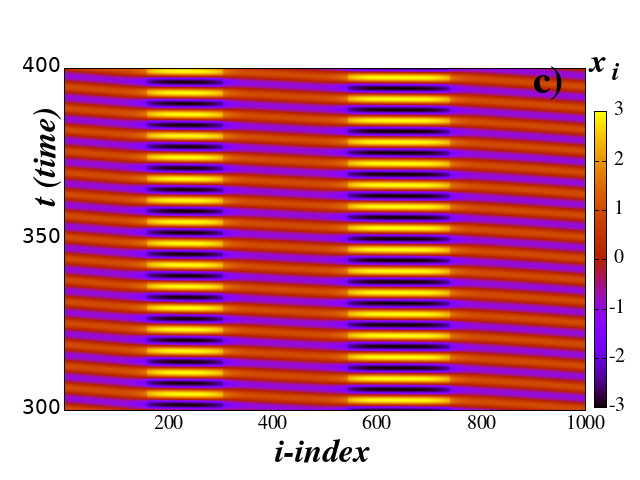}
\caption{\label{fig02} (Color online)
Two-amplitude chimera state without entanglement: 
a) Typical $x-$variable temporal evolution (elements 100 and 200 are shown), b) the $\omega$-variable profile  and c)
the space time plot of the $x_i-$variables.
Parameters are: $c_r=-1$, $r_l=1$, $r_c=2$, $r_h=3$, $N=1000$, $S=40$, $a=1.0$, $\omega =1$, $\sigma =0.3$ 
and $\sigma_{r}=0.9$. All simulations start from random initial conditions, $ r_l-1 \le r_i \le r_h+1 $.
}
\end{figure}

\par
As an example, we present in Fig.~\ref{fig02} the amplitude chimera  for the working parameter set with specific parameter
values:  $S=40$,  $\sigma =0.3$ and $\sigma_{r}=0.9$. In panel a) we present the temporal evolution of the
$x-$variables for oscillators 100 and 200, in panel b) the $\omega$-variable profile and in (c) the spacetime plot of the
$x-$variable for all elements. 
The amplitude chimera has two domains of high amplitude oscillations as shown in Fig.~\ref{fig02}c with color label
covering the range [-3,3] (black-purple-red-yellow regions) and two domains of lower amplitudes with color label ranging 
between [-1,1] (red-purple regions). All elements have the same frequency, $\omega =1$ (see Fig.~\ref{fig02}b), 
and the spacetime plot
demonstrates that there is a continuation in the spatial profiles, although the amplitude of the oscillations are
different in the different domains. This finding is typical for pure amplitude chimeras, where the different network
regions are characterized by their proper amplitudes and not by different frequencies as in the frequency chimeras.
Figure~\ref{fig02} will be used for comparison with Fig.~\ref{fig03} in the next section to demonstrate the effects
of entanglement.

\subsection{Coupled two-amplitude oscillators with entanglement}
\label{sec:coupled-with-entanglement}
\begin{figure}[H]
\centering
\includegraphics[clip,width=0.32\linewidth,angle=0]{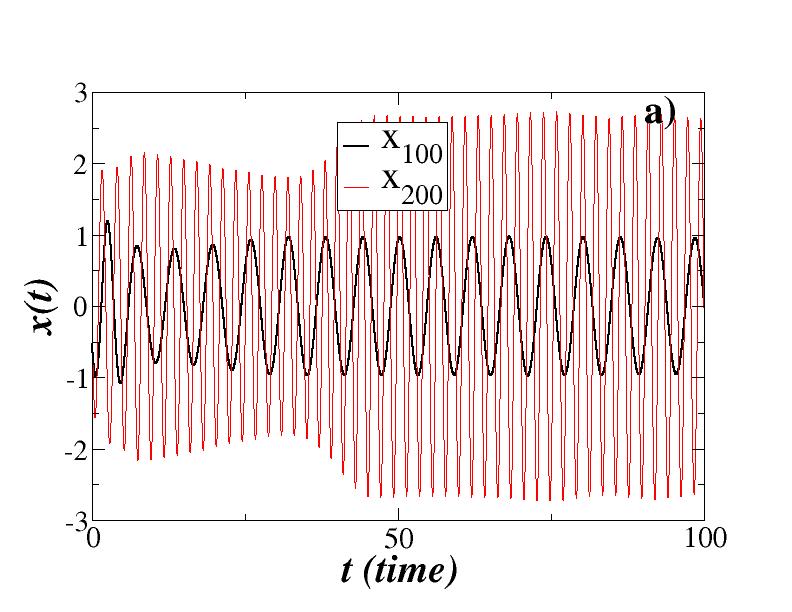}
\includegraphics[clip,width=0.32\linewidth,angle=0]{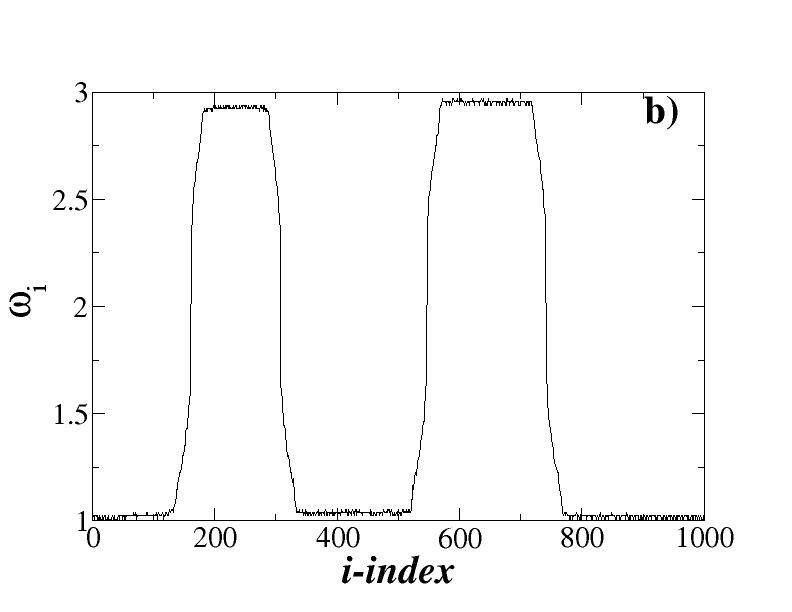}
\includegraphics[clip,width=0.32\linewidth,angle=0]{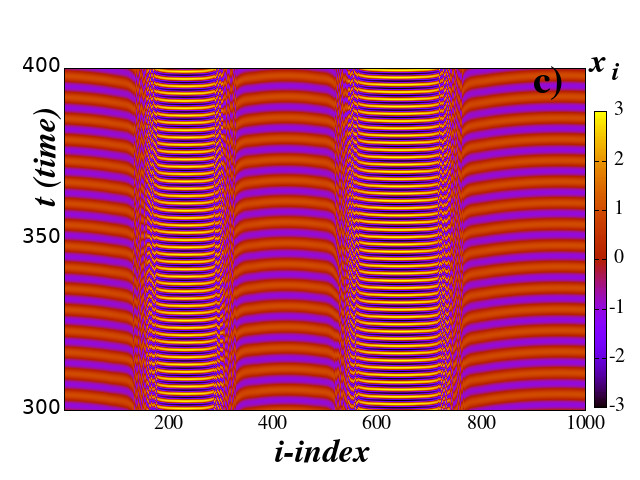}
\caption{\label{fig03} (Color online)
Two-amplitude chimera state with amplitude-frequency entanglement.
The plot in the left column depicts the temporal evolution of the $x$-variable (oscillators 100 and 200 are depicted),
the middle column the asymptotic mean phase velocity profile and the right column the spacetime plots of the $x_i$-variables: 
Parameters are: $N=1000$, $S=40$, $c_r=-1$, $r_l=1$, $r_c=2$, $r_h=3$, $a=1.0$, $\sigma =0.3$ and $\sigma_{r}=0.9$. 
All runs start from the same random initial conditions as in Fig.~\ref{fig02}.}
\end{figure}

Consider now the case of a ring network of oscillators where each element performs under frequency-amplitude entanglement.
Namely, each oscillator is linked to $S$-elements to its left and right and its frequency $\omega$ is modulated by its
amplitude $r_i$. The system of equations describing the network evolution is:
\begin{subequations}
\begin{align}
\frac{dx_i}{dt} &= -ax_i+\frac{ar_i}{\sqrt{x_i^2+y_i^2}}x_i-\boxed{\omega r_i} y_i + 
\frac{\sigma}{2S}\sum_{j=i-S}^{i+S}\left[ x_j-x_i\right]
\label{eqno09a} \\
\frac{dy_i}{dt} &= -ay_i+\frac{ar_i}{\sqrt{x_i^2+y_i^2}}y_i+\boxed{\omega r_i} x_i + 
\frac{\sigma}{2S}\sum_{j=i-S}^{i+S}\left[ y_j-y_i\right]
\label{eqno09b} \\
\frac{dr_i}{dt} &= c_r (r_i -r_l)(r_i -r_c)(r_i -r_h)+\frac{\sigma_r}{2S}\sum_{j=i-S}^{i+S}\left[ r_j-r_i\right] .
\label{eqno09c}
\end{align}
\label{eqno09}
\end{subequations}
Note the boxed expressions $\boxed{\omega r_i}$ in Eqs.~\eqref{eqno09}a and b, which account for the entanglement
of the oscillator common internal frequency $\omega$ with the individual amplitudes $r_i(t)$. All other terms in Eqs.~\eqref{eqno09}
are the same as in Eqs.~\eqref{eqno08}. Typical simulation results are depicted in Fig.~\ref{fig03}. Initial 
conditions are identical to the ones used in Fig.~\ref{fig02}.

In panel \ref{fig03}a, we present the temporal evolution of oscillators Nos. 100 and 200. We note that they have
different amplitude and different frequency. Regarding frequency, in 100 Time Units (TUs) oscillator No. 100
has performed approx. $\sim$8.5 cycles, while No. 200 has performed $\sim$25 cycles. This is verified in panel \ref{fig03}b,
where the mean phase velocity $\omega_{200}=3\cdot\omega_{100}$. We recognize here the effect of entanglement, which 
modifies the mean phase velocity in certain regions of the network, creating a symbiosis of amplitude-and-frequency
chimeras. This is further verified in the spacetime plot, panel \ref{fig03}c, where the oscillations in regions between
$\sim$[200-300] and $\sim$[600-750] have high frequencies and amplitudes, while in regions $\sim$[1-200], $\sim$[300-600]
and $\sim$[750,1000] they have low amplitude and frequency. In between the regions of constant amplitude (or frequency)
the elements have variable amplitude (or frequency) to bridge the gap between these different values.

\par As discussed in the previous section, the frequency variations in the network come as a result of the entanglement
between a variable amplitude $r$ and a constant $\omega$ characterizing the uncoupled elements. The random initial conditions
drive parts of the system to select one or the other possible amplitude and, as a consequence, they also create domains of
 different frequencies.

\subsection{Quantitative indices}
\label{sec:quantitative}

In this section we introduce the frequency and amplitude deviations as measurable indicators 
that quantify the level of network synchronization in both entangled and non-entangled scenarios.

\par To compute the limiting frequency (or mean phase velocity) $\omega_i$ of an individual oscillator $i$ in the network
we count the number of full cycles $Q_i(\Delta T )$ that the oscillator $i$ has performed 
within a relatively large time interval $\Delta T$. Then, the mean phase velocity is defined as
the number of complete cycles per unit time, namely
\begin{equation}
\omega_i = \frac{2\pi Q_i(\Delta T)}{\Delta T}.
\label{eqno10}
\end{equation}
To avoid transient phenomena, the time period $\Delta T$ used for the recording of the limiting (asymptotic) $\omega$-values
needs to be set after the system (network) has attained its steady state (i.e., after the transient period).
 
\par The above frequency definition concerns individual oscillators in the network. To quantify collective effects, 
comparative indices such as maximum and minimum mean phase velocities need to be introduced in the system.
For any given network, let us denote by $\omega_{\rm max}$ and $\omega_{\rm min}$ the maximum and
 minimum mean phase velocity recorded among all elements, respectively. Note that $\omega_{\rm max}$ and $\omega_{\rm min}$
need to be recorded after the transient period, when the network has reached the steady state.
Given that $\omega_{\rm max} \ge \omega_{\rm min} \ge 0$, let us denote $\Delta\omega =\omega_{\rm max} - \omega_{\rm min} \ge 0$. 
If $\Delta \omega =0$ (within a tolerable error $\epsilon >0$) then all oscillator in the system have the same frequency
and then frequency chimera states are not developed. Alternatively, if  $\Delta \omega > \epsilon$ the network
has the possibility to support frequency chimera states, since the characteristic of the typical (frequency) chimeras is the
non-negligible frequency difference between coherent and incoherent domains. 

\par Similarly to the definition of the frequency deviation in the typical chimera states, in the amplitude chimeras
we compute the maximum and minimum amplitudes recorded in the network. If $r_i$ denotes the limiting value of the
amplitude attained by oscillator $i$ after a long time (i.e., after the transient), 
let us denote by $r_{\rm max}=\max \{r_1, r_2,\cdots r_N \}\ge 0$
and by  $r_{\rm min}=\min \{ r_1, r_2,\cdots r_N \} \ge 0$. By their definition $r_{\rm max} \ge r_{\rm min} \ge 0$.
The amplitude divergence of the network may be quantified by the difference $\Delta r= r_{\rm max} - r_{\rm min} \ge 0$.
Similarly to the case of frequencies, if $\Delta r =0$ then all network elements are characterized by the same
amplitude and amplitude chimeras are not developed. Alternatively, if $\Delta r \ge 0$ (with an interval of toleration)
then amplitude chimeras are
possible. In the most complex case where both $\Delta r $ and $\Delta \omega $ are non-negligible, then chaotic
chimeras emerge with both amplitude and frequency variations.

\par Apart the quantitative indices $\Delta r $ and $\Delta \omega $ introduced above, a number of other indices
have been proposed in the literature, such as the Kuramoto order parameter, Fourier spectra and various correlation 
indices \cite{kuramoto:2002,krischer:2016} mainly for the study of frequency chimeras.  In the present study,
primarily $\Delta \omega $ and $\Delta r $ will be used for discrimination between frequency and amplitude chimeras.

\par In the next sections we explore the effects of various parameters in the formation of
domains of alternating amplitude/frequency values.
Without loss of generality, in the rest of the manuscript we use the following working parameter set: 
$c_r=-1$ (to ensure of the existence of one repulsive fixed point, $r_c$,
surrounded by two attractive ones $r_l$ and $r_h$), 
$r_l=1$, $r_c=2$, $r_h=3$, $R=1$, $\omega =1$ (constant frequency of all oscillators),
 $S=40$ (unless otherwise stated), $a=1.0$ and  $\sigma =0.3$.  The special case $r_l=0$  related to the presence
of bump states is discussed in Sec.~\ref{sec:results-for-bump-states}.

\section{Variations with the coupling parameters}
\label{sec:parameters}

In this section we keep all other parameters fixed to the working set and we monitor the network properties
with variation on two coupling parameters, the coupling range $S$ and coupling constant $\sigma_r$
governing the amplitude modulations. As before, we present separately the results, first without the $r-\omega$
entanglement and then with entanglement.

\subsection{Results without $r-\omega$ entanglement}
\label{sec:results-without-entanglement} 

\par To start,we here present color coded 2D maps of the maximum, minimum and difference of $r$ values formed
in the system under the working parameter set defined at the end of Sec.~\ref{sec:quantitative}, namely:
$c_r=-1$, 
$r_l=1$, $r_c=2$, $r_h=3$, $R=1$, $a=1.0$ and $\sigma =0.3$. 
The scanning parameters,  $\sigma_r$ and $s=2S/N$, are in the ranges: $0 \le \sigma_r \le 1$ and $0 \le s \le 0.2$.

\par
\begin{figure}[h]
\includegraphics[clip,width=0.32\linewidth,angle=0]{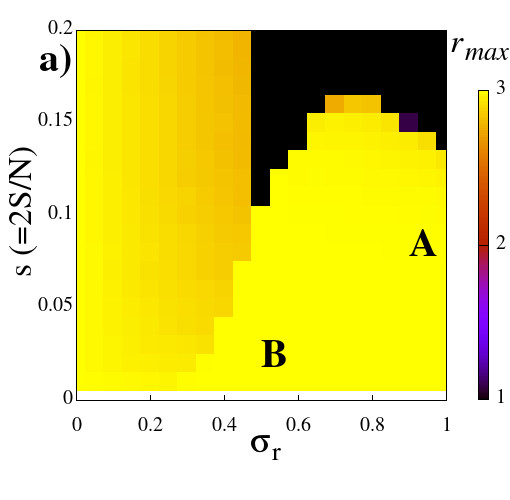}
\includegraphics[clip,width=0.32\linewidth,angle=0]{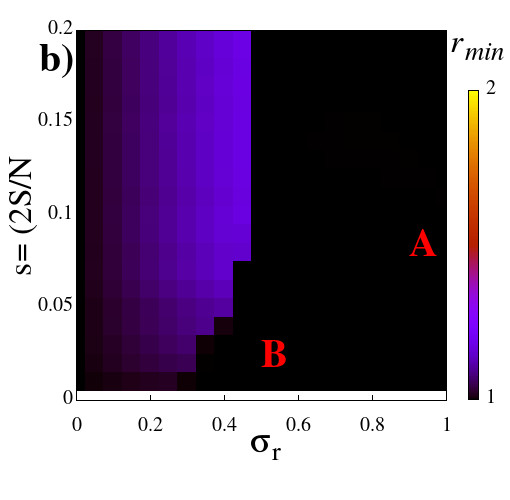}
\includegraphics[clip,width=0.32\linewidth,angle=0]{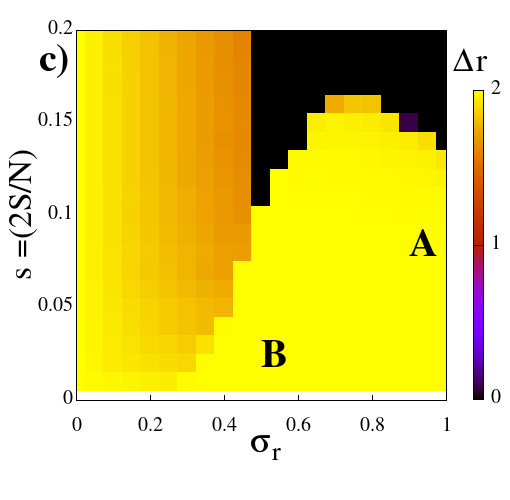}
\caption{\label{fig04} (Color online)
Color coded maps representing the values of  
a) $r_{\rm max}$, b) $r_{\rm min}$ and c) $\Delta r=r_{\rm max}-r_{\rm min}$.
Parameters are: $c_r=-1$, $r_l=1$, $r_c=2$, $r_h=3$, $N=1000$, $S=40 \> (s=0.08)$, $a=1.0$ and $\sigma =0.3$ .
 All simulations start from random initial conditions, $ r_l-1 \le r_i \le r_h+1 $.
}
\end{figure}

 Starting with the minimum $r$-values, in Fig.~\ref{fig04}b we note that the minimum values are mostly around or
slightly higher than the value $r_l$ (set to 1 in the simulations). Thus, the coupling in the system does not significantly
affect the low fixed point of the amplitudes. On the other hand, $r_{\rm max}$ changes drastically as shown in Fig.~\ref{fig04}a:
 for large values of
$\sigma_r$ and $s$ the $r_h (=3)$ amplitude fixed point is absorbed by the low amplitude fixed point $r_l =1$.
For this reason, in Fig.~\ref{fig04}c the difference $\Delta r=0$ is recorded, where the black color dominates. In these cases all elements
have the same amplitude and therefore no amplitude chimeras are possible.
 For large values of
$\sigma_r$ and low values of $s$ the $r_h (=3)$ amplitude fixed point remains close to its original values, 
while the lowest amplitudes (min values) remain close to $r_l =1$ and in this case multi-amplitude chimeras are possible.
In the corresponding regions of Fig.~\ref{fig04}c the difference $\Delta r$ takes maximum values (close to 2) and the yellow color dominates.
These regions correspond to well developed chimeras. The parameters of the example presented in Fig.~\ref{fig02} are denoted  
by the letter A in the three panels of Fig.~\ref{fig04}. One may recognize
the difference in the amplitude in panels \ref{fig02}a and c.
Finally, we note also intermediate values of the difference $\Delta r$ for intermediate values of $\sigma_{r}$.

\par
\begin{figure}[h]
\includegraphics[clip,width=0.32\linewidth,angle=0]{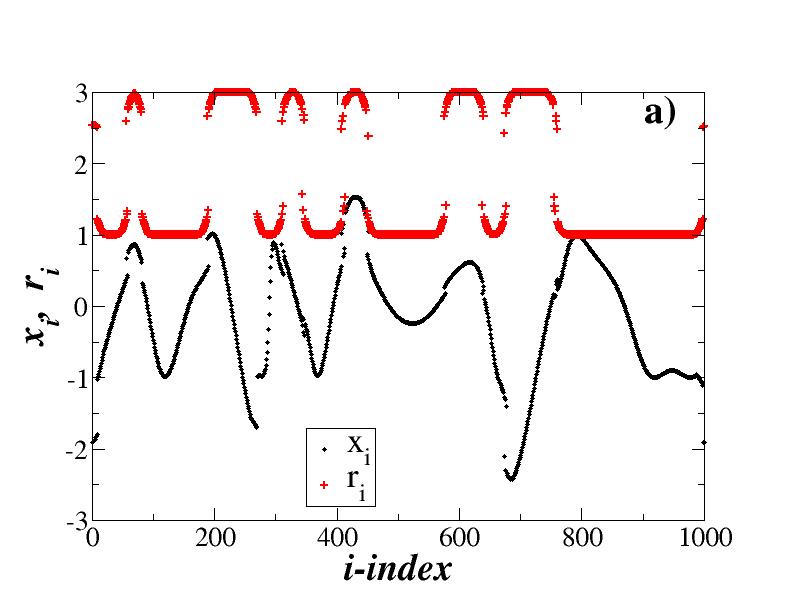}
\includegraphics[clip,width=0.32\linewidth,angle=0]{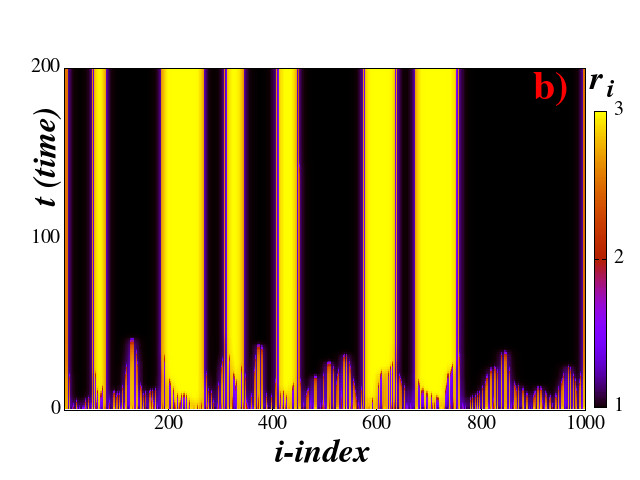}
\includegraphics[clip,width=0.32\linewidth,angle=0]{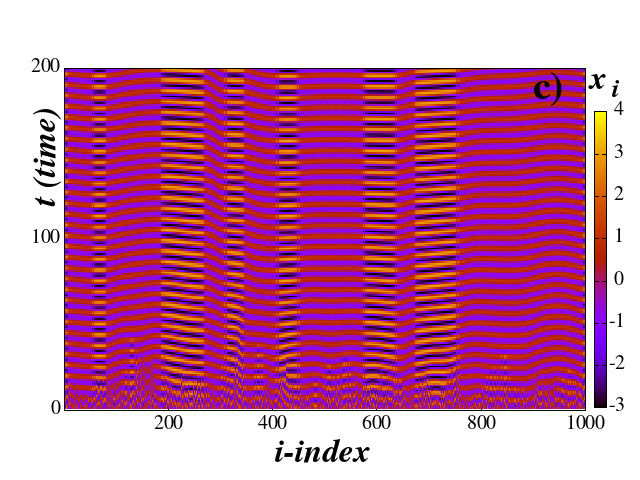}
\caption{\label{fig05} (Color online)
Amplitude multi-chimera states without entanglement.
a) The state variables $x_i$ and the amplitudes $r_{i}$ at 200 TU, b) Spacetime plot of the system amplitudes 
and c) Spacetime plots of the variables $x_i$, $i=1\cdots N$.
Parameters are: $c_r=-1$, $r_l=1$, $r_c=2$, $r_h=3$, $N=1000$, $S=10 \> (s=0.02)$, $a=1.0$, $\sigma_r=0.5$ and $\sigma =0.3$ .
 All simulations start from random initial conditions, $ r_l-1 \le r_i \le r_h+1 $.
}
\end{figure}

 As a second indicative example, we present in Fig.~\ref{fig05} the case of parameters $s=0.02$ and $\sigma_r=0.5$, which are denoted
by the letter B in Figs.~\ref{fig04}. The reduction in the values of the coupling range $s$ causes the split of the network in many regions of
alternating high and low amplitude dynamics. This inverse relation between the coupling range and the number of distinct domains
is also known to take place in other dynamical systems coupled in networks \cite{omelchenko:2013,tsigkri:2016}.
To be more precise, in Fig.~\ref{fig05}a we present a typical snapshot of the system's state variables $x_i$ at 200 TU (black dots) together
 with the system amplitudes $r_i$ (red crosses).   It is easily understood that the system splits into domains of low amplitude (close to $r_l=1$)
and high amplitudes $r_h=3$. In fact, Fig.~\ref{fig05}b demonstrates that the oscillators reach their final amplitudes as early as 30TUs and
maintain these values afterwards. For shorter times ($t< 30$ TU) many regions of different amplitudes
are present, as imposed by the random initial conditions. As time advances, and with the influence of the coupling terms $\sigma_r$ in
the system, adjacent regions merge and at the final/steady state six regions of high (and six of low) amplitudes
prevail. A discontinuous change of the variable $r_i$ and $x_i$ can be observed in the borders between the twelve different
domains and this is confirmed by the spacetime plots in Fig.~\ref{fig05}c. In this respect, and in view of Fig.~\ref{fig05}b
these states can be considered as amplitude death states.

\par
\begin{figure}[h]
\begin{center}
\includegraphics[clip,width=0.5\linewidth,angle=0]{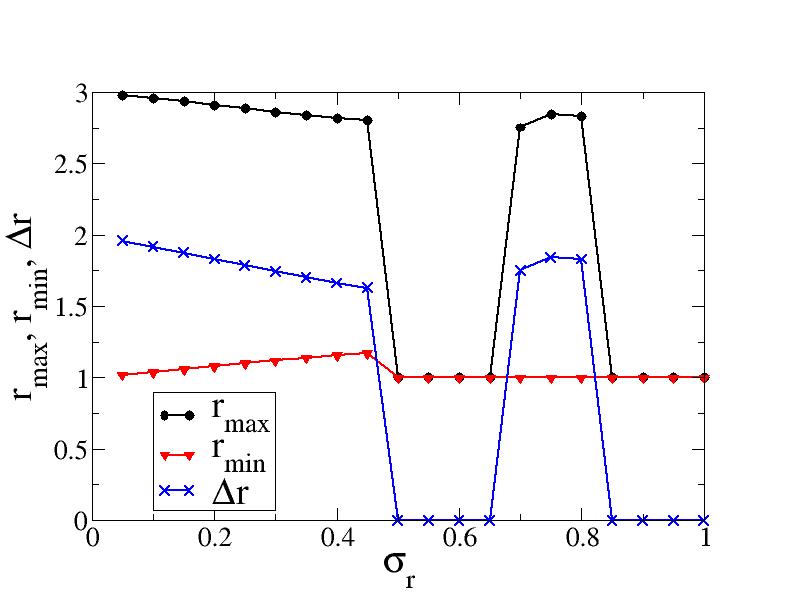}
\end{center}
\caption{\label{fig06} (Color online)
Comparative representation of the measures  
 $r_{\rm max}$ (black circles), $r_{\rm min} $ (red triangles) and $\Delta r=r_{\rm max}-r_{\rm min}$ (blue X-symbols).
Parameters are as in Fig.~\ref{fig04} and $s=0.16$. All simulations start from random initial conditions, $ r_l-1 \le r_i \le r_h+1 $.
}
\end{figure}

 To get a better comparison of the measures $r_{\rm max}$, $r_{\rm min}$ and $\Delta r$ we plot in the same graph the three measures
as a function of $\sigma_{r}$ for constant $s=0.16$. The results are shown in Fig.~\ref{fig06}. In the left part of the graph,
where $\sigma_r < 0.45$, the maximum amplitude in the system stays close to $r_h=3$ with a tendency to decrease as the coupling range
increases. Indeed, as $\sigma_r$ increases the interactions in the system lower the amplitude to approach lower values, closer to $r_l$.
Similarly, the minimum amplitude in the system stays close to $r_l=1$ with a tendency to increase as the coupling strength
increases. As $\sigma_r$ increases the interactions in the system increase the lower amplitudes to approach the maximum values, close to $r_h$.
Around a critical value $\sigma_r\sim 0.5$, the amplitudes collapse to the lowest value $r_{\rm max}\sim r_{\rm min} \sim 1$ and $\Delta r=0$.
This behavior persists for the interval $0.5 \le \sigma_r \le 0.7$ where all amplitudes are equal in the system and amplitude chimeras are
not possible. Increasing further the $\sigma_r$ values, in the interval $0.7 \le \sigma_r \le 0.85$, a divergence of amplitudes develops in the
network and the maximum amplitudes reach values close (but lower) than $r_h$, while the minimum values stay near $r_l$ indicating a region
where amplitude chimeras are present. Finally, for  $\sigma_r > 0.85$ $r_{\rm max}= r_{\rm min}$ and amplitude chimeras disappear.
\par To show that in this case we have no significant difference in the frequencies (mean phase velocities) of the elements, we present the color
coded maps of $\omega_{\rm max}$ and $\Delta \omega$. The map for $\omega_{\rm min}$ is not shown because it is very similar to $\omega_{\rm max}$ since $\Delta \omega$'s
are mostly 0 in the system.
\begin{figure}[h]
\begin{center}
\includegraphics[clip,width=0.32\linewidth,angle=0]{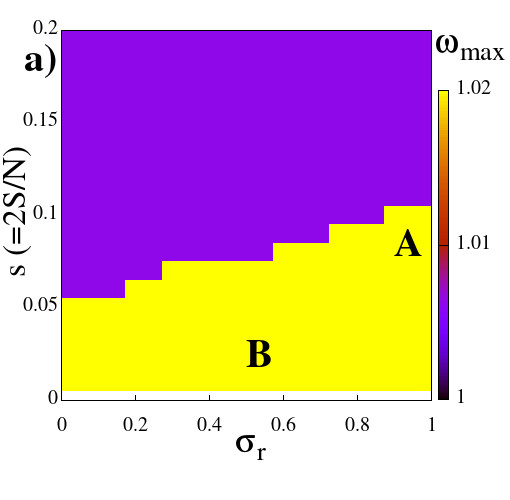}
\hspace{0.8cm}
\includegraphics[clip,width=0.32\linewidth,angle=0]{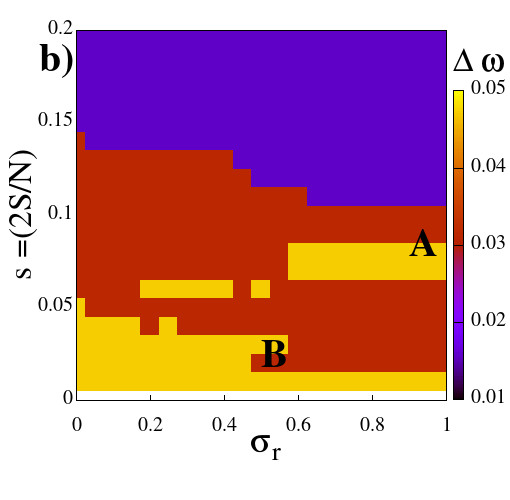}
\end{center}
\caption{\label{fig07} (Color online)
Color coded maps representing the values of  
a) $\omega_{\rm max}$ and b) $\Delta \omega=\omega_{\rm max}-\omega_{\rm min}$.
Parameters are as in Fig.~\ref{fig04}.
 All simulations start from random initial conditions, $ r_l-1 \le r_i \le r_h+1 $.
}
\end{figure}
\par From Fig.~\ref{fig07}a we note a change in the colors of the color map representing the $\omega_{\rm max}$ values, but the change is negligible 
as can be seen from the $\omega$-scale shown on the right of the panel. The same conclusions are drawn from panel b, representing the $\omega$-differences
in the system. Therefore, frequency chimeras are not supported in this system without entanglement and only amplitude chimeras are observed
in the parameter regions discussed above.

\subsection{Results with $r-\omega$ entanglement}
\label{sec:results-with-entanglement} 

\par We now turn to the case where entanglement is introduced in the system as discussed in Sec.~\ref{sec:coupled-with-entanglement} and this
induces frequency chimeras in addition to amplitude ones. The parameters are the same as in Fig.~\ref{fig03} and in this section we explore
the region of parameters were frequency chimeras occur. The scanning parameters are again in the range $0 \le \sigma_r \le 1$ and $0 \le s=2S/N \le 0.2$.

\par
\begin{figure}[h]
\includegraphics[clip,width=0.32\linewidth,angle=0]{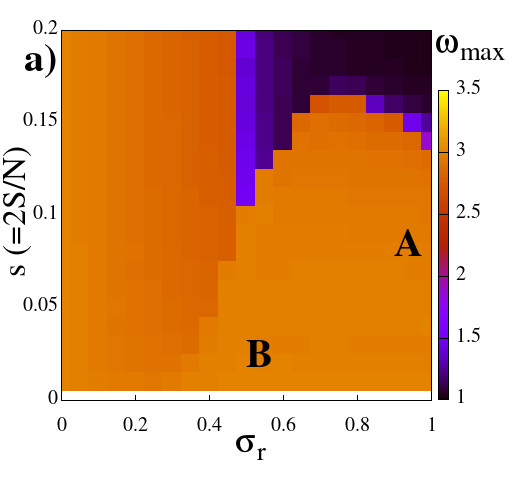}
\includegraphics[clip,width=0.32\linewidth,angle=0]{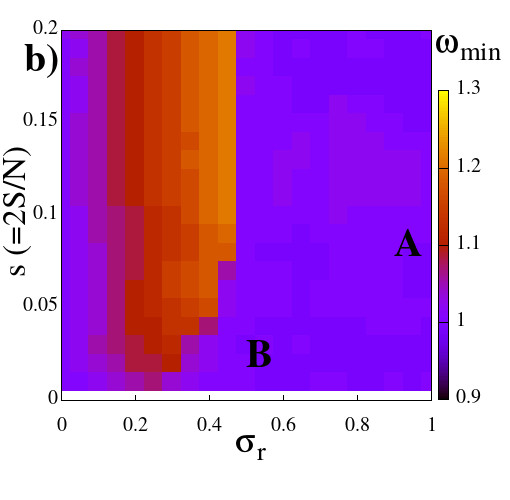}
\includegraphics[clip,width=0.32\linewidth,angle=0]{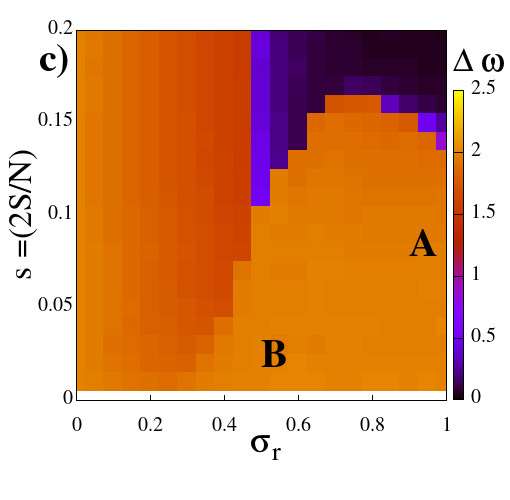}
\caption{\label{fig08} (Color online)
Color coded maps representing the values of  
a)  $\omega_{\rm max}$, b) $\omega_{\rm min}$ and c) $\Delta \omega=\omega_{\rm max}-\omega_{\rm min}$.
Parameters are as in Fig.~\ref{fig03}
 All simulations start from random initial conditions, $ r_l-1 \le r_i \le r_h+1 $.
}
\end{figure}

 
In Fig.~\ref{fig08}, we use color coded maps to present only the variations
 $\omega_{\rm max}$, $\omega_{\rm min}$ and $\Delta \omega=\omega_{\rm max}-\omega_{\rm min}$. We do not need to show
the maps for the $r_{\rm max}$, $r_{\rm min}$ and $\Delta r$ because they are identical with the ones in Fig.~\ref{fig04}. This is because the
amplitudes are formed only via Eq.~\eqref{eqno09c} which evolves independently of $\omega$ and is identical to Eq.~\eqref{eqno08c}. These last two equations
develop independently of Eq.~\eqref{eqno09a} and Eq.~\eqref{eqno09b} (or Eq.~\eqref{eqno08a} Eq.~\eqref{eqno08b} for the non-entangled case). 
In the simulations, while the values of $r$ are formed via Eq.~\eqref{eqno09c} (or Eq.~\eqref{eqno08c} for the non-entangled case) they are fed in
Eqs.~\eqref{eqno09a} and ~\eqref{eqno09b} and they contribute to the formation of the $\omega$-profiles. Note, that the mechanisms of $r$-feeding
in the entangled and non-entangled case are different as the boxed terms in Eq.~\eqref{eqno08a} and Eq.~\eqref{eqno09a} demonstrate, 
and that is why in the first case the $\omega$-profiles are vaguely modified, while in the
second case they vary considerably following closely the amplitude variations, see Fig.~\ref{fig08}. Note in panel \eqref{eqno08c} the $\Delta \omega$ variations
which are in the range $0 \le \Delta \omega \le 2.5$ and compare with the non-entangled case, Fig.~\ref{fig07}b, where the variations are negligible, 
$0 \le \Delta \omega \le 0.02$.

\par
\begin{figure}[h]
\includegraphics[clip,width=0.32\linewidth,angle=0]{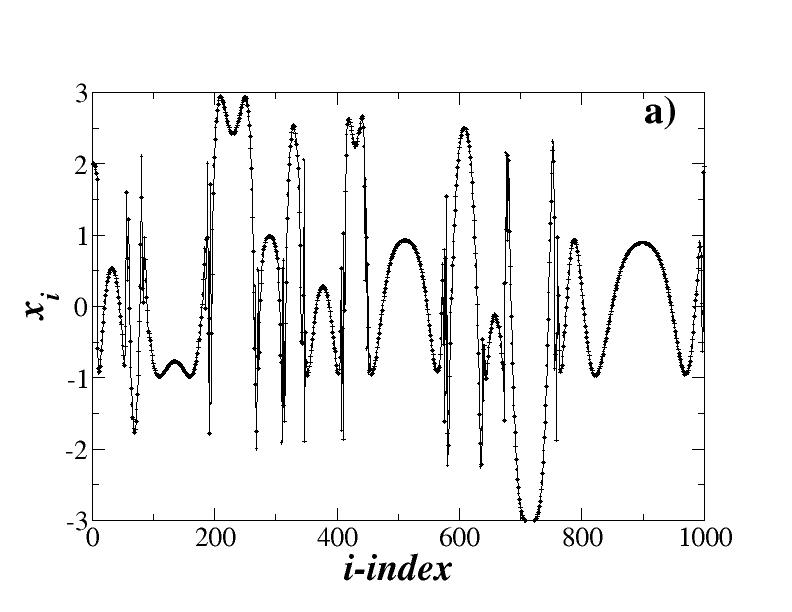}
\includegraphics[clip,width=0.32\linewidth,angle=0]{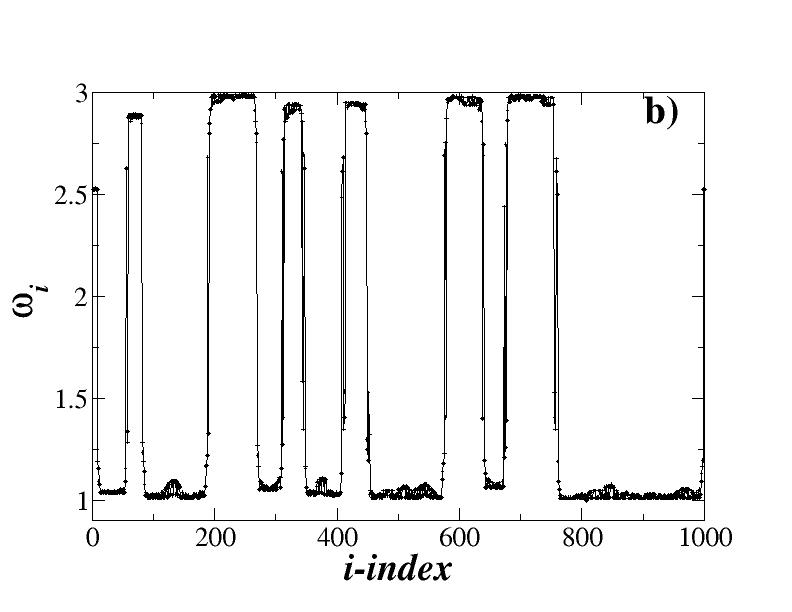}
\includegraphics[clip,width=0.32\linewidth,angle=0]{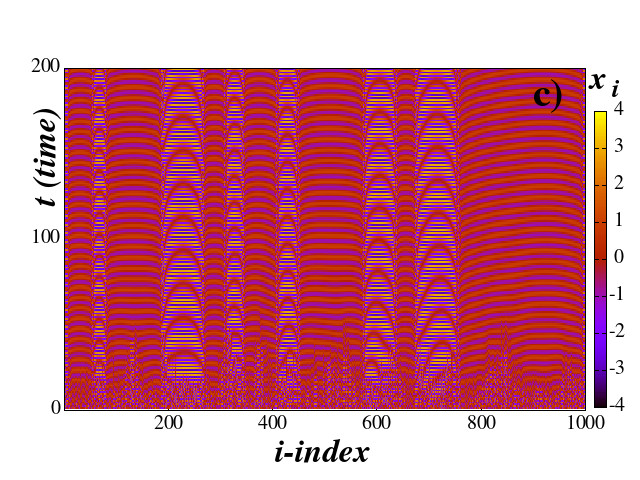}
\caption{\label{fig09} (Color online)
Amplitude multi-chimera states with entanglement.
a) A typical profile of the state variables $x_i$ at 200 TU, b) The $\omega_i$ profile after 800 TUs 
and c) Spacetime plots of the variables $x_i$, $i=1\cdots N$.
Parameters are: $c_r=-1$, $r_l=1$, $r_c=2$, $r_h=3$, $N=1000$, $S=10 \> (s=0.02)$, $a=1.0$, $\sigma_r=0.5$ and $\sigma =0.3$ .
 All simulations start from random initial conditions, $ r_l-1 \le r_i \le r_h+1 $.
}
\end{figure}

 In parallel with the discussions in the previous section we also investigate the synchronization properties 
at parameter position denoted by letter B on maps Fig.~\ref{fig08}, in the case of entanglement. 
In Fig.~\ref{fig09}a we show a typical profile of the state variable $x_i$. Many coherent  regions are formed separated by small incoherent ones 
as may be observed in this figure. We recall that the amplitude ($r_i$) profile is identical to the one in Fig.~\ref{fig05}b since they are governed 
by the same equations (see Eq.~\eqref{eqno08}c and Eq.~\eqref{eqno09}c). On the other hand the mean phase velocity profiles are different
as can be seen in Fig.~\ref{fig09}b, because it is governed by the entanglement terms (see Eqs.~\eqref{eqno09}a and b) and as a result
a number of successive domains with binary frequencies are observed. The presence of these domains becomes clear in the spacetime plots, see
Fig.~\ref{fig09}c. In contrast to the case of no entanglement, at the borders between coherent regions of different frequencies one may
observed small incoherent domains. The effect here (Fig.~\ref{fig09}b and Fig.~\ref{fig09}c) is vaguely observable due to the small size
(large number) of the coherent regions and is not as clear as 
in Fig.~\ref{fig03}b and Fig.~\ref{fig03}c result, where only a few (four) coherent regions are observed with larger sizes.

\par
\begin{figure}[h]
\begin{center}
\includegraphics[clip,width=0.5\linewidth,angle=0]{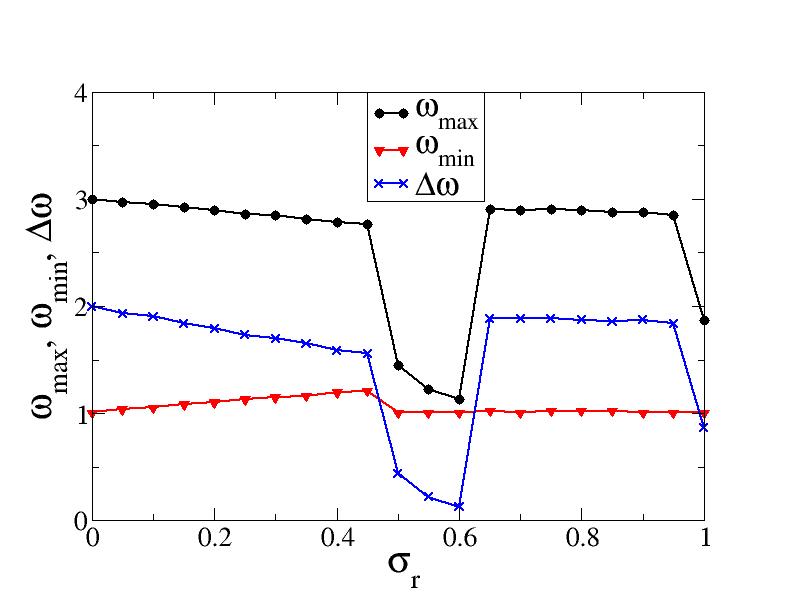}
\end{center}
\caption{\label{fig10} (Color online)
Comparative representation of the measures  
 $\omega_{\rm max}$ (black circles), $\omega_{\rm min} $ (red triangles) and $\Delta \omega=\omega_{\rm max}-\omega_{\rm min}$ (blue X-symbols).
Parameters are as in Fig.~\ref{fig04} and $s=0.14$. All simulations start from random initial conditions, $ r_l-1 \le r_i \le r_h+1 $.
}
\end{figure}

 To have a visual idea of the frequency variations in the system, we also present the values of $\omega_{\rm max}$, $\omega_{\rm min}$ and $\Delta \omega$
for constant $s=0.14$. The results are shown in plot Fig.~\ref{fig10}. The form of the three curves follow closely the ones in Fig.~\ref{fig06}, as the $r$-variations
induce the $\omega$ ones. Namely, for low $\sigma _r$ values the 
maximum frequency in the system stays close to the product $r_h \cdot \omega =3\cdot 1=3$ with a tendency to decrease as the coupling strength
increases. In the middle range of parameter $\sigma_r$ there is an abrupt drop and the maximum and minimum frequencies take close values,
while in the large $\sigma_r$ range the maximum frequency increases again close to the product $r_h \cdot \omega =3$.
The minimum frequency in the system stays always close to the product $r_l \cdot \omega =1\cdot 1=1$ and varies slightly above it. 
As a result, the frequency divergence $\Delta \omega$ remains close to 0 in the middle of $\sigma_r$ range and frequency chimeras are not supported
in this area, while for $\sigma_r$ values outside this range frequency chimeras are possible, as demonstrated in Fig.~\ref{fig03}.
We also note a tendency for the maximum and minimum frequency to approach each other as $\sigma_r\to 1$, similarly to the tendency of $r_{\rm min}$
and $r_{\rm max}$ in the case without entanglement, see Fig.~\ref{fig06}. (Note that Fig.~\ref{fig06} and Fig.~\ref{fig10} have been recorded
for different $s$ values, 0.16 and 0.14 respectively).

\par As a final remark related to effects of entanglement,
we would like to recall that in all case (with or without entanglement) the parameter $\omega$, which primarily governs the oscillators frequency, is kept constant.
Only in the case of entanglement the frequencies of the different oscillators are carried along with the amplitude $r$ variations. In the case of no entanglement the frequency of all
oscillators remains close to the constant $\omega$ value. Amplitude variations leading to amplitude chimeras take place in both cases in the present model.

\par We would like to add that it is possible to form a more complex extension of these models with four equations, to explicitly incorporate both amplitude and
frequency variations. From the point of view of the dynamical system, this means to add also Eq.~\eqref{eqno01}c in Eqs.~\eqref{eqno03} and we end up with a system 
of four equations and four variables: $x,\> y, \> r, \> \omega$. 
In this case one may combine different frequencies and amplitudes in the system which, together with the coupling terms, will
induce even more complex synchronization patterns. Furthermore, the four variable model ($x,\> y, \> r, \> \omega$) may be considered with and without entanglement.
These extensions fall outside the scope of the present investigation and are left for future studies.

\section{Bump states: The case of a fixed point vanishing to 0}
\label{sec:results-for-bump-states} 

\par We now investigate the scenario in which one of the fixed points takes the value 0. Because we have assumed that $0\le r_l \le r_c \le r_h$,
we can now set the lowest fixed point $r_l=0$. In the  case of a single uncoupled oscillator this means that depending on the initial conditions, 
the oscillator, if attracted by $r_l$, will cease
to oscillate and will remain quiescent. To the contrary, if the (single) oscillator is attracted onto fixed point $r_h$,  it will oscillate with amplitude $r_h$.
In the case of coupled elements in a network, we expect that depending on their initial states 
some of the oscillators will fall on one or the other fixed point  and we may observe
coexistence of active and quiescent oscillators. 
\par In this section we first investigate the case of $r_l=0$ without entanglement, in Sec.~\ref{sec:fix-point-null-noentenglement} and in Sec.~\ref{sec:fix-point-null-entenglement} we report on the case of $r_l=0$ with entanglement. Different type of bump states are observed in each case: a) In the case of no-entanglement
quiescent regions are observed which coexist with active regions oscillating with constant frequency but with different amplitudes. b) In the case of entanglement
the quiescent regions coexist with active regions oscillating with variable frequencies and amplitudes. As will be discussed in the sequel,
in  case b) regions of coherent
and incoherent motion can be observed in the system due to the frequency variability.

\subsection{A vanishing fixed point in the case of no entanglement}
\label{sec:fix-point-null-noentenglement}

\begin{figure}[H]
\centering
\includegraphics[clip,width=0.32\linewidth,angle=0]{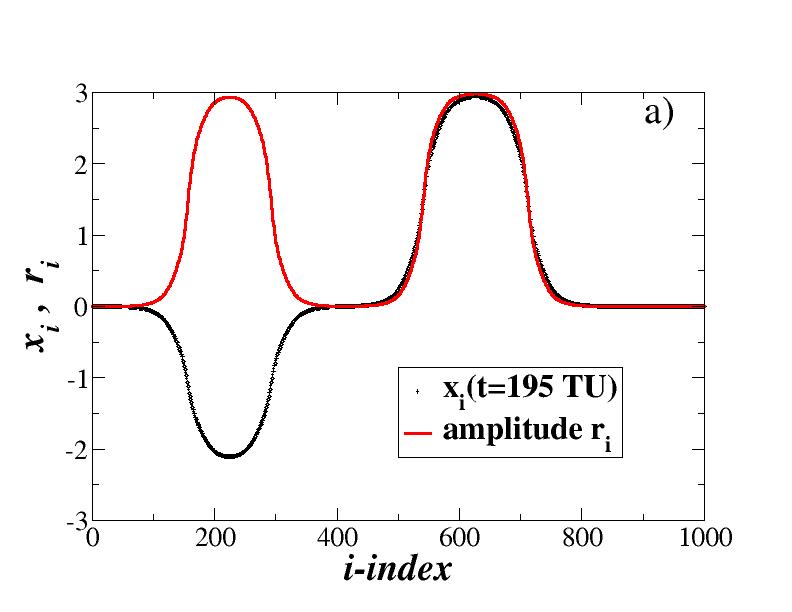}
\includegraphics[clip,width=0.32\linewidth,angle=0]{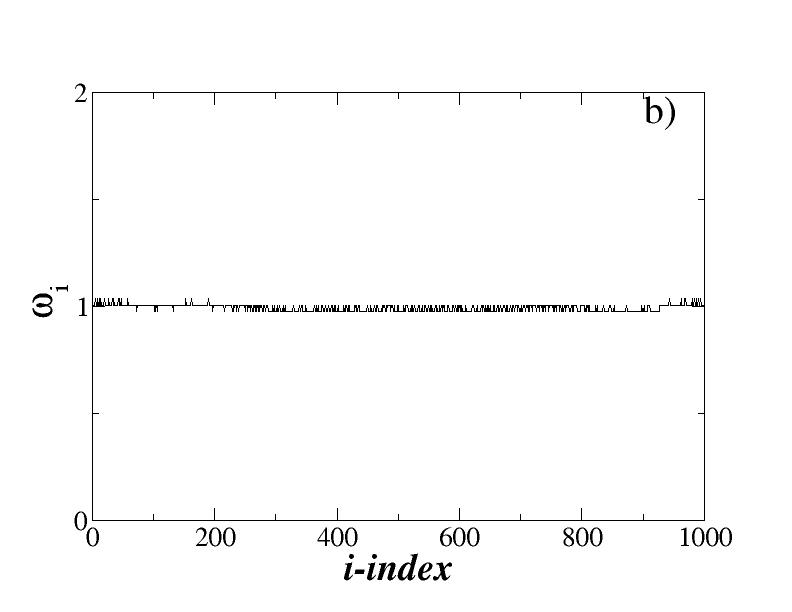}
\includegraphics[clip,width=0.32\linewidth,angle=0]{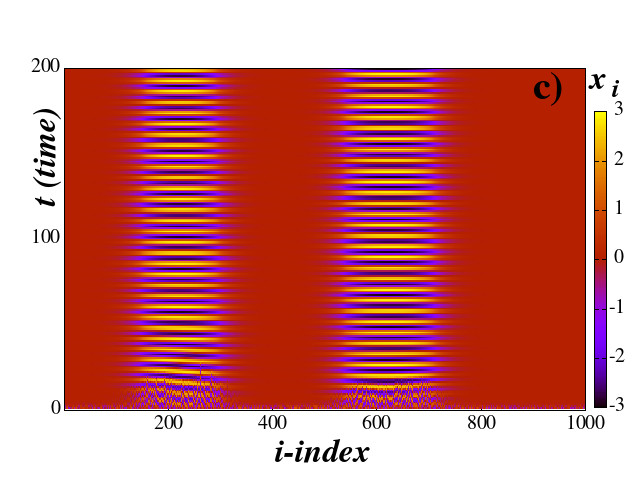}
\caption{\label{fig11} (Color online)
Amplitude bump state {\bf without} amplitude-frequency entanglement and with $r_l=0$.
The plot in the left column depicts a typical $x$-variable profile (time $t=195$ TUs, black line) and the amplitudes $r_i$ (red line),
the middle column the mean phase velocities and the right column the spacetime plots of the $x_i$-variables: 
Parameters are: $N=1000$, $S=40$, $c_r=-1$, $r_l=0$, $r_c=1.5$, $r_h=3$, $a=1.0$, $\omega =1$, $\sigma =0.5$ and $\sigma_{r}=3.5$. 
All runs start from the same random initial conditions as in Fig.~\ref{fig02}.}
\end{figure}

Starting with the case of no entanglement and $r_l=0$, we present in Fig.~\ref{fig11}a a snapshot of the $x_i$-variables of the ring nodes in black color.
All nodes follow each other in synchrony, however, they do not attain a common maximum amplitude as the red curve in  Fig.~\ref{fig11}a indicates.
This red curve presents the maximum amplitude that each element reaches. One clearly understands that there are two 
quiescent regions Q$_1 \sim$ [0-100, 800-1000] and Q$_2 \sim$ [400-500]
where the maximum amplitude is 0 (note that due to the ring connectivity with periodic boundary conditions
the elements 0-100 and 800-1000 belong to the same quiescent region, Q$_1$). This means that the elements in regions Q$_1$ and Q$_2$ do not oscillate in reality
 but they remain inactive fluctuating around the $r_l=0$ fixed point. The elements in the intermediate regions [100-400] and [500-800] perform synchronous
oscillations of variable continuous amplitudes as the red curve in Fig.~\ref{fig11}a shows. The synchrony in the system is demonstrated in Fig.~\ref{fig11}b,
where all frequencies recorded are basically the same, $\omega_i=1$ for $i=1,\cdots 1000$. 

\par The spacetime plot in Fig.~\ref{fig11}c gives a clearer picture
of the temporal evolution in the system. In fact, the regions Q$_1$ and Q$_2$ do not change in space and time and they remain constantly
near $x_i=0$. It is interesting to compare Fig.~\ref{fig02}c and ~\ref{fig11}c : in the former case regions Q$_1$ and Q$_2$ perform oscillations 
with lower amplitudes
than the intermediate regions ([100-400] and [500-800]), while in the latter case regions Q$_1$ and Q$_2$ are quiescent.
The two intermediate regions in Fig.~\ref{fig11}c, [100-400] and [500-800], present oscillations in time, with their state variables $x_i$
 taking antipodal values. This can
be better observed in \ref{fig11}a, where the elements of region [100-400] are in the negative values at the same time that the ones in region [500-800]
are on the positive axis. 

\par We stress here, that these are the amplitude bump states. Namely, the quiescent regions oscillate with null amplitudes while the active
regions coexist with the quiescent ones and they present a variety of amplitudes, as shown in Fig.~\ref{fig11}a (red points). The amplitudes
of the active regions may vary in the interval $(r_l=0,r_h]$. The interval is open on the left side because the 0-amplitude characterizes the
quiescent regions, while in the active regions all elements have amplitudes above 0 and they may potentially reach values as large as $r_h$.

\subsection{A vanishing fixed point in the case of entanglement}
\label{sec:fix-point-null-entenglement}

\begin{figure}[H]
\centering
\includegraphics[clip,width=0.32\linewidth,angle=0]{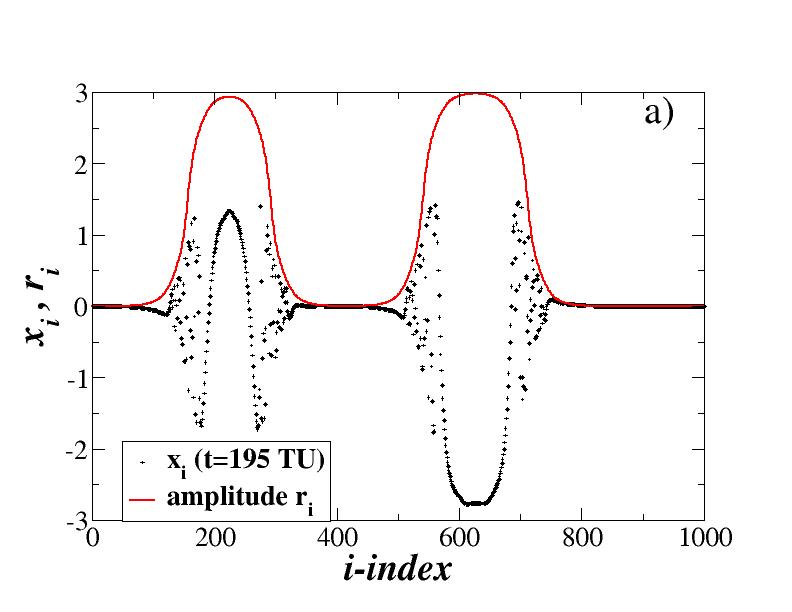}
\includegraphics[clip,width=0.32\linewidth,angle=0]{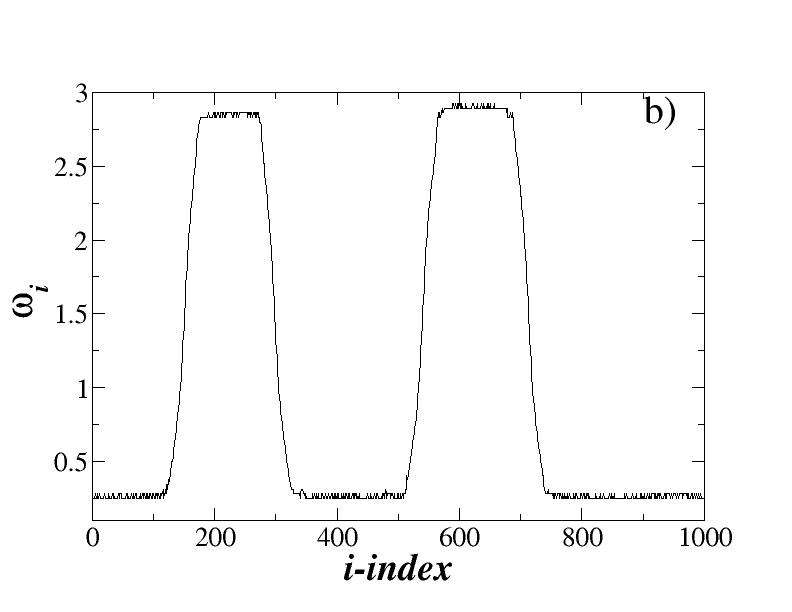}
\includegraphics[clip,width=0.32\linewidth,angle=0]{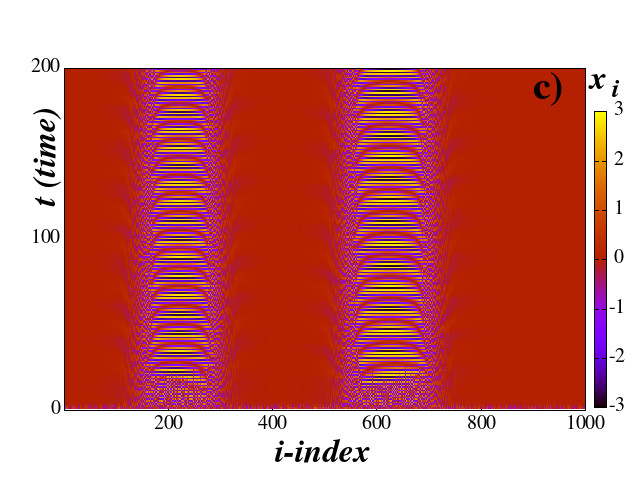}
\caption{\label{fig12} (Color online)
Amplitude-frequency bump state {\bf with} entanglement and $r_l=0$.
The plot in the left column depicts a typical $x$-variable profile (time $t=195$ TUs, black dots) and the amplitudes $r_i$ (red line),
the middle column the mean phase velocities and the right column the spacetime plots of the $x_i$-variables: 
Parameters are: $N=1000$, $S=40$, $c_r=-1$, $r_l=0$, $r_c=1.5$, $r_h=3$, $a=1.0$, $\sigma =0.5$ and $\sigma_{r}=3.5$. 
All runs start from the same random initial conditions as in Fig.~\ref{fig02}.}
\end{figure}

In the case of amplitude-frequency entanglement the system evolution changes drastically. The elements loose synchrony because the amplitude variations
influence the node frequencies. That is the reason why some nearby elements move incoherently, as can be seen in Fig.~\ref{fig12}a, where the black dots indicate
the positions of the system's state variables in a typical snapshot. 
\par To be more precise, in Fig.\ref{fig12}a we observe a typical system snapshot at $t=195$TU (black dots) together with the maximum amplitude $r_i$ attained by each node. As previously, regions Q$_1$ and Q$_2$ are quiescent and the state variables $x_i$ stay near 0. At the intermediate regions there are a number of elements that are incoherent (at the borders with regions Q$_1$ and Q$_2$) while in the middle of these regions the elements become coherent again. The reason for this behavior is understood if we inspect the mean phase velocities $\omega_i$, in Fig.\ref{fig12}b. Due to the amplitude-frequency entanglement the intermediate regions acquire a variation of frequencies ranging from a minimum $r_l \cdot \omega =0$ (in regions Q$_1$ and Q$_2$) to a maximum $r_h \cdot \omega =3\cdot 1=3$
(in the intermediate regions). We note, that due to the nonlinearity of the system equations
together with the coupling the maximum $\omega_i$ values never reach as high as 3. To bridge the gap between the quiescent elements and the ones oscillating with frequencies close to $\omega_{\rm max}$, the mediating elements (regions [120-180], [270-325], [510-560] and [700-750]) have variable frequencies and therefore their $x_i$ state variables are incoherent. The different regions can be also observed in Fig.~\ref{fig12}c, where the quiescent regions are colored red, the regions of 
constant frequency close to $\omega_{\rm max}$ show the oscillations while the regions between oscillatory and quiescent elements have mixed colors indicating 
incoherence. As a last remark, we stress the comparison between images Figs.~\ref{fig03}c and ~\ref{fig12}c. Both cases present similar oscillations in the
high (close to $\omega_{\rm max}$) frequencies. However, in regions Q$_1$ and Q$_2$ 
the vanishing of the fixed point amplitude $r_l=0$ together with the amplitude-frequency
entanglement causes the cease of oscillations in Fig.~\ref{fig12}c, while low frequency oscillations are maintained in Fig.~\ref{fig03}c.

\par In the case of entanglement frequency-and-amplitude bump states are produced. This is because the amplitude drives the frequency and the active regions
present variability both in amplitude and in frequency. To the contrary, the elements that belong to quiescent regions are inactive and their amplitudes
as well as frequencies  register nearly null.

\section{Conclusions}
\label{sec:conclusions}
\par Motivated by the observation of amplitude chimeras, notably in coupled chaotic dynamical systems, we propose here a toy model which is able to
produce amplitude chimeras when coupled in a network. This becomes possible via a two-amplitude bifurcation mechanism.
This toy model is a modification of a previous one \cite{provata:2020}, which was designed to specifically produce frequency chimeras.

\par The present toy model was further extended to include  amplitude-frequency entanglement terms. The modified model gives rise to complex amplitude-and-frequency
chimeras, as the amplitude variations drive the frequency through the entanglement terms. Apart the amplitude and complex chimeras this model can
produce bump states when one of the fixed points of the uncoupled system is set to zero. This way the non-entangled model produces amplitude bump
states, while combined amplitude together with frequency bump states are possible when entanglement is considered in the network.

\par The scenarios presented here can serve as a general mechanism for formation of amplitude chimeras and bump states. 
As in the case of the toy model of Ref.~\cite{provata:2020} leading to frequency chimeras, we may argue that even in cases 
where the mechanisms of producing amplitude chimeras or bump states are not explicit, the combination of coupling together 
with the nonlinearity in the dynamics may generate bistability (or multistability)
in the oscillator amplitude, indirectly, giving rise to multi-amplitude states.

\par In the present study the influence of the coupling parameters $\sigma_r$ and $s$ was considered. It would be interesting to investigate
the influence of the other parameters $(c_r , \sigma, \alpha)$, as well as multiple fixed points leading to amplitudes states with many levels.
Along the same research lines, in the system dynamics one may include equations that allow for distinct bifurcations in amplitude and frequency, 
creating this way
more complex synchronization schemes, as was described at the end of Sec.~\ref{sec:results-with-entanglement}.  
Further relevant studies may be directed toward investigations of amplitude chimeras and bump states in 2D and 3D systems and in multiplex networks.
Also, the search for other solvable toy models can lead to alternative mechanisms inducing complex synchronization phenomena.

\par As a final note we may propose to investigate the case of non-exponential relaxation. In Ref.~\cite{provata:2018} except for the case of 
the exponential relaxation to the limiting periodic trajectory, power law relaxation has been proposed.
Exploring whether this power law relaxation dynamics also provokes the emergence of amplitude and/or frequency chimeras and/or bump states
would be a subject worth of further consideration.


\section*{Acknowledgements}

\par The author would like to thank K. Anesiadis and Dr. J. Hizanidis for helpful discussions.
This work was supported by computational time granted from the Greek Research \& Technology Network (GRNET) 
in the National HPC facility - ARIS - under Project IDs  PR012015  and PR014004.

\section*{Data availability statement}
The data that support the findings of this study are available from the corresponding author upon reasonable request.
\section*{Conflict of interest}
The author declares no conflicts of interest.
\bibliographystyle{iopart-num}
\providecommand{\newblock}{}


\end{document}